# Rejoinder: Monitoring Networked Applications With Incremental Quantile Estimation

John M. Chambers, David A. James, Diane Lambert and Scott Vander Wiel

## 1. DIVERSITY OF MONITORING GOALS AND CONSTRAINTS

There are many kinds of networks, each with many types of variables and monitoring goals. Our paper addressed only one of the countless possible combinations of network and monitoring goals. We are grateful to the discussants for expanding our paper by providing insights into other network monitoring problems that present different challenges to statisticians.

Denby, Landwehr and Meloche (DLM) describe three network monitoring problems, each with different requirements for detection speed, communication constraints and scalability. The Voice over Internet protocol (VoIP) application, for example, requires good scalability, low overhead and quick responses to problems that manifest in a variety of quality-of-service (QoS) metrics. Monitoring service-level agreements, on the other hand, needs a prompt signal when path transit times become too long—a more focused goal than the VoIP problem. Our monitoring problem is most similar to DLM's third example, monitoring call centers through flexible reporting of historical reliability and performance. These problems typically have a wide variety of analytic goals, some of which are not determined until an analyst begins to drill through high-level summaries into data slices that show unusual behavior.

Whereas DLM concentrate on full-path QoS for VoIP, Lawrence, Michailidis and Nair (LMN) describe a QoS problem in which path measurements are used to estimate link-level characteristics, presumably for the purpose of managing the network, perhaps by modifying routing tables, adding key links or upgrading hardware at nodes.

To the list of monitoring problems that we and the discussants have described, we would add detection of worm outbreaks (Bu, Chen, Vander Wiel and Woo, 2006), dynamic thresholding of error counts (Lambert and Liu, 2006), fraud detection (Cahill, Lambert, Pinheiro and Sun, 2002) and call blocking events (Becker, Clark and Lambert, 1998). And there are certainly others that we are overlooking.

The variety of applications raised by the reviewers and our own experience demonstrate that there is no canonical statistical problem in the domain of monitoring networks for performance and reliability. In our application, the software architects imposed a hard constraint that the summary records had to have a fixed length and would be transmitted at regular intervals. Also, the requirement for a very small footprint stemmed from the need for the agent software to run on personal computers that may be old and slow and may be connected to the network by a low bandwidth link. While the quantile estimates must be reasonably accurate, the growth plan for the business placed much more emphasis on ease of implementation for new features and upgraded architecture to improve scalability. Therefore, improvements to quantile accuracy had to be made with relatively low development (software coding) cost. The simplicity of Incremental Quantiles (IQ) was obviously attractive.

## 2. DATA COMPRESSION

DLM, LMN and Yu all discuss connections that the IQ algorithm has to methods for compressing and sketching data streams. Although compression was not likely to be used in our application, it is critical for sensor networks, for example, where data transmission is much more costly. We hope that Yu and others will pursue statistical compression methods that allow updating summaries without decompression.

## 3. SMOOTHING AND DETECTION PERFORMANCE

LMN advocate that, for monitoring purposes, "the procedure should be devised to estimate the current scenario" and then outline how exponentially weighted moving averages (EWMAs) could be formed using either quantiles or cumulative distribution functions (CDFs).

We like the idea of extending IQ to compute EWMAs of CDFs and, in fact, we proposed this possibility to the product managers of the monitoring software. However, they were not prepared to modify the meaning of the basic summaries computed by agents. One reason for their reluctance is that temporal changes in performance characteristics represent just one type of anomaly that analysts want to uncover. Other anomalies are topographically defined. For example, an outage might affect only a small group of users over an extended period of time. Furthermore, appropriate EWMA weight parameters will differ according to the goals of the analyst, and these goals could vary widely. Therefore EWMA calculations would need to be done in real time at the server in our application and not by the agents.

Yu outlines a scheme that would track the *current* CDF using a moving window of data, processed in blocks that are small enough for within-block stationarity to be a reasonable assumption. A moving window of blocks would not be difficult to implement, although EWMAs would achieve much the same goal with less complexity because an EWMA scheme would use only the previous quantile estimates and the new data in **D** and would have the same level of complexity as the nominal IQ algorithm.

DLM, LMN and Yu all were dissatisfied that we did not explore performance of the monitoring scheme in terms of false alarm rates and detection times. Although we agree that good detection performance is, in general, an important design goal, the portion of the software suite that uses IQ does not attempt to produce real-time alarms of anomalous events; that aspect of monitoring is handled by a companion system that analyzes network event data. Nevertheless, the procedure that DLM sketch in which an agent emits a summary record when triggered by a low $p$-value for testing the hypothesis of a change in distribution is a reasonable approach to the on-line detection problem if changes are large enough to be detected by individual agents. The problem is more difficult, however, if the signal for a problem is buried in noisy data and distributed over many agents. In this case, two-way communication between the agents and the server could be valuable. Furthermore, if the goal is dynamic response to an emerging problem, then the information being shared will need to extend beyond evidence of a change and include the character of the change as well.

## 4. ACCURACY AND EFFICIENCY

LMN explain that the computational cost of IQ is $\mathcal{O}(N \log(N))$ or even up to $\mathcal{O}(N^2)$. It is important to clarify that $N$ is the fixed length of the **D**-buffer and therefore the sorting operation represents a fixed amount of overhead for each round of the IQ algorithm. IQ is linear in terms of the total number of data elements that are processed through the algorithm. The computational complexity of sorting comes into play when considering the price of improving the accuracy by growing **D**, but in practice modern sorting algorithms are extremely efficient even for large, but memory-resident, blocks of data.

LMN discuss $\varepsilon$-approximate algorithms that appear in the computer science literature. These guarantee that an estimate is within $\varepsilon$ of the correct quantile level; for example, $\varepsilon = 0.01$ assures that the $p = 0.98$ quantile estimate lies between the actual 0.97 and 0.99 sample quantiles. Accuracy that is uniform in $p$ is appropriate for constructing approximate equidepth histograms but tail quantiles need high $p$-resolution that seems difficult to achieve with $\varepsilon$-approximate algorithms. We would like to see the $\varepsilon$-approximate algorithms extended to provide accuracy that improves in the tails. For example, if an algorithm reports the $q$th sample quantile as an estimate of the $p$th sample quantile, then we would like a guarantee that the logit values of $p$ and $q$ differ by less than $\varepsilon$. IQ has no such guarantee, but neither does any other algorithm, as far as we are aware.

All the discussants have raised problems that remain to be addressed. We thank them and the Editor for helping to raise awareness of the many statistical issues that remain to be resolved in the context of network monitoring.

## REFERENCES

BECKER, R., CLARK, L. and LAMBERT, D. (1998). Events defined by duration and severity, with an application to net-work reliability (with discussion). *Technometrics* **40** 177–194.

Bu, T., Chen, A., Vander Wiel, S. and Woo, T. (2006). Design and evaluation of a fast and robust worm detection algorithm. In *Proc. IEEE INFOCOM 2006*. IEEE Press, Piscataway, NJ.

Cahill, M., Lambert, D., Pinheiro, J. and Sun, D. (2002). Detecting fraud in the real world. In *Handbook of Massive Datasets* 911–929. Kluwer, Dordrecht.

Lambert, D. and Liu, C. (2006). Adaptive thresholds: Monitoring streams of network counts. *J. Amer. Statist. Assoc.* **101** 78–88. MR2252435